\documentclass[aps,prd,twocolumn,preprintnumbers,superscriptaddress,nofootinbib]{revtex4-2}
\pdfoutput=1

\usepackage[height=8.85in,width=6.45in]{geometry}

\usepackage[utf8]{inputenc}
\usepackage[T1]{fontenc}
\usepackage{amsmath,amssymb,amsthm,amsfonts,mathrsfs,mathtools}
\usepackage{ragged2e}
\usepackage{enumitem}
\usepackage{young,bm,braket}
\usepackage{dsfont}
\usepackage[vcentermath]{youngtab}
\usepackage{tikz}
\usetikzlibrary{shapes}
\usetikzlibrary{calc}
\usetikzlibrary{arrows.meta}
\usetikzlibrary{positioning}
\usetikzlibrary{decorations.markings}
\usetikzlibrary{decorations.pathmorphing}
\usetikzlibrary{fadings}

\allowdisplaybreaks
\usepackage{slashed}

 \usepackage{hyperref}
 \hypersetup{
 	pdfstartview={FitH},    
 	pdftitle={From Integrability to Chaos in Tensionless AdS$_3$ String Theory},    
 	pdfauthor={Matthias R.~Gaberdiel, Beat Nairz, Cheng Peng},     
 }

\usepackage{graphicx}
\usepackage{tikz}

\usetikzlibrary{decorations.markings}
\tikzset{line/.style={line width=0.25mm},
	curve/.style={line,smooth,tension=1},
	->-/.style={decoration={
			markings,
			mark=at position #1 with {\arrow[>=stealth]{>}}},postaction={decorate}},
	-<-/.style={decoration={
			markings,
			mark=at position #1 with {\arrow[>=stealth]{<}}},postaction={decorate}},
}

\usepackage{times}
\usepackage[scaled]{couriers}

\usepackage{xcolor}
\usepackage{mdframed}

\renewenvironment{figure}[1][]{
	\begin{originalfigure}[#1]
		\begin{mdframed}[linecolor=black!0,backgroundcolor=black!0]
		}{
		\end{mdframed}
	\end{originalfigure}
}

\definecolor{MM1}{rgb}{0.368417, 0.506779, 0.709798}
\definecolor{MM2}{rgb}{0.880722, 0.611041, 0.142051}
\definecolor{MM3}{rgb}{0.560181, 0.691569, 0.194885}

\newcommand{\be}{\begin{equation}}
\newcommand{\ee}{\end{equation}}
\def\bal#1\eal{\begin{align}#1\end{align}}

\def\p{\pi}

\usepackage{comment}

\DeclareMathAlphabet{\mathpzc}{OT1}{pzc}{m}{it}

\newcommand{\mrm}[1]{\mathrm{#1}}
\newcommand{\mcl}[1]{\mathcal{#1}}
\newcommand{\mbb}[1]{\mathbb{#1}}

\newcommand{\mfr}[1]{\mathfrak{#1}}

\newcommand{\cint}[1]{\underset{#1}{\oint}}
\newcommand{\llangle}{\left\langle}
\newcommand{\rrangle}{\right\rangle}
\newcommand{\br}[1]{\left({#1}\right)}
\newcommand{\bbr}[1]{\big({#1}\big)}
\newcommand{\corr}[1]{\left\langle{#1}\right\rangle}

\newcommand{\ccontraction}[2]{
\begin{tikzpicture}[baseline=-.5ex]
\node[](n1)at(0,0){$#1$};
\node[](n2)[right=-1ex of n1]{$#2$};
\draw[]let \p1 = (n1),\p2=(n2) in ([yshift=-2ex]\p1)-- ([yshift=-3ex]\p1)--(\x2,\y1-3ex)node[midway]{$\times$}--(\x2,\y1-2ex);
\end{tikzpicture}
}

\newcommand{\sscontraction}[2]{
\begin{tikzpicture}[baseline=-.5ex]
\node[](n1)at(0,0){$#1$};
\node[](n2)[right=-1ex of n1]{$#2$};
\draw[]let \p1 = (n1),\p2=(n2) in ([yshift=2ex]\p1)-- ([yshift=3ex]\p1)--(\x2,\y1+3ex)--(\x2,\y1+2ex);
\end{tikzpicture}
}

\ExplSyntaxOn
\cs_new:Npn \pd {\partial}
\cs_new:Npn \ed {\mathrm d}
\cs_new:Npn \im#1 {\mathrm{im}\,{#1}}
\cs_new:Npn \Ric {\mathcal R\mathrm{ic}}
\cs_new:Npn \tr#1{
\str_if_eq:nnTF{#1}{R}
{\mathrm{tr}\c_math_subscript_token g\mathcal R\mathrm{ic}}
{\str_if_eq:nnTF{#1}{K}
{\mathrm{tr}\c_math_subscript_token h\mathcal K}
{\mathrm{tr}{#1}}
}
}
\cs_new:Npn \K {\mathscr K}
\cs_new:Npn \m {1,\cdots,m}
\cs_new:Npn \up#1#2{
\bool_if:nTF{\str_if_eq_p:nn{#1}{V} || \str_if_eq_p:nn{#1}{M}||\str_if_eq_p:nn{#1}{P}|| \str_if_eq_p:nn{#1}{O}}
{
{\mathscr #1}^{[#2]}
}
{
\bool_if:nTF{\str_if_eq_p:nn{#1}{K}||\str_if_eq_p:nn{#1}{R}||\str_if_eq_p:nn{#1}{G}||\str_if_eq_p:nn{#1}{T}}
{
{\mathcal #1}^{[#2]}
}
{
{#1}^{[#2]}
}
}
}
\cs_new:Npn \GN {G\c_math_subscript_token {\text N}}
\cs_new:Npn \diag {\mathrm{diag}}
\cs_new:Npn \Span {\mathrm{Span}}
\cs_new:Npn \SF {\mathsf{SF}}
\cs_new:Npn \II {\mathrm{I\!I}}
\cs_new:Npn \AdS {\text{AdS}}
\cs_new:Npn \ads#1 {\text{AdS}_#1}
\cs_new:Npn \Sch#1 {
\mathsf{Sch}\big[{#1}^{[i]}\big]
}
\ExplSyntaxOff

\begin{document}

\title{Non-planar corrections in the symmetric orbifold}

\author{Matthias R.~Gaberdiel}
\affiliation{Institut f\"ur Theoretische Physik,
ETH Z\"urich,
Wolfgang-Pauli-Strasse 27,
8093 Z\"urich, Switzerland}
\affiliation{Kavli Institute for Theoretical Sciences (KITS) and School of Quantum, University of the Chinese Academy of Sciences, Beijing 100190, China}

\author{Beat Nairz}
\affiliation{Institut f\"ur Theoretische Physik,
ETH Z\"urich,
Wolfgang-Pauli-Strasse 27,
8093 Z\"urich, Switzerland}

\author{Cheng Peng}
\affiliation{Kavli Institute for Theoretical Sciences (KITS) and School of Quantum,
University of the Chinese Academy of Sciences, Beijing 100190, China}
\affiliation{Peng Huanwu Center for Fundamental Theory, Hefei, Anhui 230026, China}

\begin{abstract}
We calculate the non-planar corrections to the anomalous dimensions of certain quarter BPS states in the symmetric product orbifold $\text{Sym}^N\bbr{\mbb{T}^4}$. We find that some of the degeneracies in the spectrum for large twist $w$ and large $N$ are lifted by these contributions. This suggests that integrability in the symmetric orbifold is broken by the non-planar (i.e.\  $\frac{1}{N}$) corrections.
\end{abstract}

\maketitle

\section{Introduction}

The symmetric product orbifold of $\mbb{T}^4$, denoted by $\text{Sym}^N\bbr{\mbb{T}^4}$, is the CFT dual of tensionless string theory on $\text{AdS}_3\times {\rm S}^3\times \mbb{T}^4$ with one unit of NS-NS flux \cite{Gaberdiel:2018rqv,Eberhardt:2018ouy,Eberhardt:2019ywk}. At this point in moduli space, the string theory can be described by a free field worldsheet theory \cite{Dei:2020zui} and is thus under good analytical control; in particular, this therefore allows one to prove the duality for this specific case. However, the background is rather special in that no R-R flux is present, and it is important to understand how the situation is modified once R-R flux is switched on. This is most conveniently described in the dual symmetric orbifold theory where it corresponds to perturbing the theory by an exactly marginal operator.\footnote{While R-R flux is obviously quantised, the relevant perturbation parameter is $g_s Q_{\rm R}$, and this becomes a continuous parameter in the weakly coupled ($g_s\rightarrow 0$) string description.} Thus backgrounds with R-R flux can be studied using the well-understood framework of conformal perturbation theory \cite{Lunin:2002fw,Gomis:2002qi,Gava:2002xb}, see also
\cite{Gaberdiel:2015uca,Keller:2019suk,Benjamin:2021zkn} for general aspects of these calculations.

In \cite{Gaberdiel:2023lco}, two of us studied the deformed symmetric orbifold CFT systematically.\footnote{In \cite{Gaberdiel:2023lco} this analysis was performed for the singlet perturbation; the triplet case was studied, using similar methods, in \cite{Gaberdiel:2025smz}.} As the CFT has $\mcl{N}=(4,4)$ supersymmetry, one can understand the perturbed spectrum of the theory by calculating the action of the supercharges in perturbation theory \cite{Gava:2002xb}. More specifically, in \cite{Gaberdiel:2023lco} the planar limit  ($N\to\infty$) was considered, and the anomalous conformal dimensions for large twist ($w\to\infty$) were studied. This is the regime that makes contact with the BMN limit \cite{Berenstein:2002jq} and that can be, intuitively, compared to long spin-chains in $\mcl{N}=4$ SYM. It was shown in  \cite{Gaberdiel:2023lco} that, in this limit, the situation simplifies drastically: the deformed action of the supercharges can be described by a centrally extended `off-shell' symmetry algebra, similar to the dynamic spin chain description in $\mcl{N}=4$ SYM \cite{Beisert:2005tm}. In particular, the system possesses an integrable $S$-matrix, satisfying the Yang-Baxter equations, and is thus `integrable'. Among other things this predicts additional degeneracies in the large twist spectrum, and this is indeed what was found. The degeneracy of the spectrum is lifted at finite twist $w$, and the $1/w$ corrections to the infinite twist result are quite complicated \cite{Gaberdiel:2024nge}. Finite size effects of this kind are likely to involve `wrapping' corrections, see e.g.~\cite{Ambjorn:2005wa}, and in planar ${\cal N}=4$ these lead to deviations in the spectrum once the interaction range exceeds the length of the spin chain. Because $\text{AdS}_3/\text{CFT}_2$ has flat directions, it is believed that these wrapping corrections appear already to lowest order in the perturbation, irrespective of the length of the twist sector $w$ \cite{Abbott:2015pps}, and that they can only be switched off by going to the asymptotic $w\rightarrow \infty$ limit.\footnote{We thank Juan Maldacena for a discussion about this point.}
\smallskip

In addition to the finite twist corrections, there are also non-planar contributions to the spectrum. In $\mcl{N}=4$ SYM, these contributions are known to partially lift the degeneracies of the planar limit, which is seen as an indication of the breakdown of integrability \cite{Beisert:2003tq}. Similarly, a lifting of the planar degeneracies in the $\mathfrak{su}(2)\times
\mathfrak{su}(2)$ sector of ABJM was found in \cite{Kristjansen:2008ib}. It is the aim of this paper to study the analogous effects for the symmetric orbifold. More specifically, we shall concentrate on certain quarter BPS states that have degenerate spectra in the large twist and large $N$ (planar) limit. We will calculate the $\frac{1}{N}$ corrections to their anomalous conformal dimensions and present strong evidence that they lift the degeneracy, even at large twist. The breakdown of integrability is also expected to be visible from the level spacing statistics of the spectrum \cite{Haake:2010fgh}, and this aspect of the perturbation will be studied in some detail in the companion paper \cite{GNP2}. The findings of that paper also support our assertion that these $\frac{1}{N}$ corrections indeed break integrability.

\smallskip

The paper is organised as follows. In Section~\ref{sec:non-planar_corr}, we discuss the set-up and explain how the non-planar corrections arise. Section~\ref{sec:results} contains our numerical results which show that the non-planar corrections lead to a lifting of the degeneracies. We describe the calculation in more detail in Section~\ref{sec:degeneracy_breaking} and discuss qualitative differences between the planar and non-planar calculations.
Finally, we conclude in Section~\ref{sec:conclusion}. There are two appendices where some of the more technical details are explained.

\section{Non-planar correlators}\label{sec:non-planar_corr}

In this section we briefly introduce the set-up of the orbifold calculations and explain how to determine the non-planar corrections. We will largely follow the conventions of \cite{Gaberdiel:2023lco,Gaberdiel:2024nge,Gaberdiel:2024dfw}. In the situation we are considering, the orbifold CFT is perturbed by
\begin{equation}
g\int d^2 x\ \Phi(x,\bar{x})\ ,
\end{equation}
where
\begin{equation}
\ket{\Phi} =
\frac{i}{\sqrt{2}}\,\bigl(G^-_{-\frac{1}{2}}\tilde{G}'^-_{-\frac{1}{2}}-
G'^-_{-\frac{1}{2}}\tilde{G}^-_{-\frac{1}{2}}
\bigr)\ket{2}\ .
\end{equation}
Here, the state $\ket{2}$ is the $h=j=\frac{1}{2}$ BPS state in the 2-cycle twisted sector, making $\Phi$ exactly marginal. Its structure as a supercharge descendant of a BPS state furthermore guarantees that this deformation preserves the $\mcl{N}=(4,4)$ supersymmetry algebra \cite{Dixon:1987bg}.

\subsection{Multi-cycle intermediate states}

We are interested in determining the anomalous dimensions of various quarter BPS states in the perturbed orbifold. The relevant mixing-matrix $\tilde{\gamma}$ (whose eignvalues are the anomalous dimensions) can be calculated as a perturbed two-point function, but we shall make use of the algebraic relation
\begin{equation}
\tilde{\gamma} = \{\tilde{S}_2,\tilde{Q}_2\} = \tilde{L}_0 - \tilde{K}^3_0\ ,
\end{equation}
where we have defined
\begin{equation}
\tilde{S}_2 = \tilde{G}'^+_{-\frac{1}{2}}\ ,\qquad \tilde{Q}_2 = \tilde{G}'^-_{\frac{1}{2}}\ ,
\end{equation}
see Appendix \ref{app:conventions} for our conventions. Thus, $\tilde{\gamma}$ can also be written as a sum over products of three-point functions \cite{Keller:2019suk},
\begin{equation}\label{eq:gamma_decomposition}
\begin{aligned}
\bra{\phi'}\tilde{\gamma}\ket{\phi}
={}& \sum_{\chi}\Bigl[
\bra{\phi'}\tilde{S}_2\ket{\chi}
\bra{\chi}\tilde{Q}_2\ket{\phi} \\
&\quad+
\bra{\phi'}\tilde{Q}_2\ket{\chi}
\bra{\chi}\tilde{S}_2\ket{\phi}\Bigr]\ .
\end{aligned}
\end{equation}
The advantage of writing $\tilde{\gamma}$ in this way is that each three-point function is easy to calculate; the disadvantage is that the number of possible intermediate states $\ket{\chi}$ is very large \cite{Gaberdiel:2024nge}.

To calculate $\tilde{\gamma}$, we thus have to evaluate the action of a right-moving supercharge
$\tilde{S}_2=\tilde{G}'^+_{-1/2}$, say, on the quarter BPS states $\ket{\phi}$ in the $w$-twisted sector. These states are dual to single-particle states in string theory \cite{Eberhardt:2018ouy}. In the undeformed theory, $\tilde{S}_2$ annihilates these states, but there can be transitions at first order in the coupling $g$. However, because $\Phi$ is in the 2-cycle twisted sector, $\tilde{S}_2$ must map the state into a different twisted sector. Which twisted sectors are allowed is controlled by the $S_N$ group structure of the orbifold \cite{Pakman:2009zz}. Let us denote by $\sigma_\rho$ the operator which implements the twist by $\rho$. In order to make the operator gauge-invariant, we need to average over all group elements in the conjugacy class $[\rho]$ of $\rho$, i.e.\ we need to consider
\begin{equation}\label{average}
\sigma_{[\rho]} \propto \sum_{g\in S_N} \sigma_{g \rho g^{-1}}\ .
\end{equation}
Then the OPE of two such gauge-invariant operators takes the form
\begin{equation}\label{OPE}
\sigma_{[\rho]}\cdot \sigma_{[\tau]} \sim \sum_{g,h\in S_N} \sigma_{g \rho g^{-1}h \tau h^{-1}}\ .
\end{equation}
The right-hand-side can again be written in terms of averages of the form (\ref{average}), but in general various conjugacy classes will appear.

For the case at hand, we have a quarter BPS state in the $w$-cycle twisted sector, i.e.\ a descendant of the single-cycle twist operator
\begin{equation}
\sigma_w := \sigma_{[(1\cdots w)]}\ ,
\end{equation}
while $\Phi$ is associated to the $2$-cycle twisted sector. The OPE of $\Phi$ with $\sigma_w$ will thus contain contributions from the conjugacy classes $[\rho]$ that appear on the right-hand-side of  eq.~(\ref{OPE}).
The leading contributions in the planar (i.e.\ large $N$) limit come from the terms where $\sigma_{[\rho]}$ is again a single-cycle twist operator, and there are two possible cases (up to relabelling), namely
\begin{align}
(1\,w)\cdot (1\cdots w) &= (1\cdots w-1)\ ,\nonumber\\
(1\,w+1)\cdot (1\cdots w) &= (1\cdots w+1)\ .
\end{align}
To leading order in $1/N$, the perturbation by $\Phi$ therefore maps the $w$-cycle twisted sector to either the $(w-1)$- or $(w+1)$-cycle twisted sector, and correspondingly, these are the sectors for $\ket{\chi}$ that contribute to $\bra{\chi}\tilde{S}_2\ket{\phi}$ in the planar limit.
At large $N$ they contribute to leading order since the corresponding correlators scale as \cite{Pakman:2009zz}
\begin{equation}\label{planarcor}
\corr{\sigma_{w \pm 1}\,\sigma_2\,\sigma_w} \sim \frac{1}{\sqrt{N}}\ .
\end{equation}
Geometrically, these contributions arise from covering maps whose covering surface is a sphere; these are the terms that were considered in \cite{Gaberdiel:2023lco}.

The subleading contributions at large $N$ come from the terms where the $2$-cycle mixes two colours that are not adjacent in the $w$-cycle. (They are therefore `non-planar' also in this sense!) This leads to conjugacy classes $\rho$ that are of multi-cycle form,
\begin{equation}
\begin{aligned}
(1\,k+1)\cdot (1\cdots w)
&= (1\cdots k)(k+1\cdots w)\ ,\\
&\hspace{-2em} k=2,\dots,w-2\ .
\end{aligned}
\end{equation}
We denote the corresponding twist operator by $\sigma_{(w-k,k)}$, and the associated three-point function is then $\frac{1}{\sqrt{N}}$ suppressed with respect to the planar correlator (\ref{planarcor}) above,
\begin{equation}
\corr{\sigma_{(w-k,k)}\,\sigma_2\,\sigma_w} \sim \frac{1}{N}\ .
\end{equation}
This is the non-planar\footnote{The diagram corresponding to this three-point function is again planar, i.e.~the covering surface is a sphere. However, it computes an OPE channel associated with a torus diagram of the perturbed two-point function, and thus gives rise to a subleading (in $\frac{1}{N}$) contribution to the anomalous dimension.} correction that we shall be considering in this paper. Note that the remaining possible transitions are disconnected correlators, where the 2-cycle and $w$-cycle do not overlap, and they do not contribute to the anomalous dimension.

\begin{figure}[]
\centering
\includegraphics[width=\columnwidth]{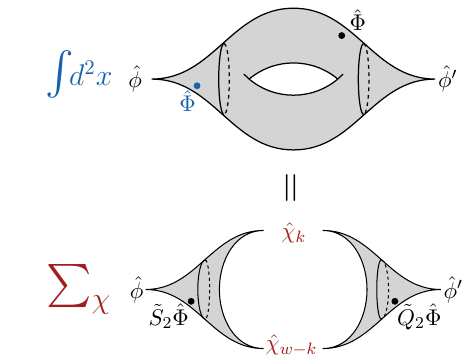}
\caption{A sketch of the non-planar contributions to the anomalous mixing matrix $\bra{\phi'}\tilde{\gamma}\ket{\phi}$ under the perturbation $\Phi$. The non-planar corrections can be calculated either by integrating over possible torus-coverings of the perturbed two-point function (above), or by considering the perturbed action of the supercharges $\tilde{S}_2,\tilde{Q}_2$ with a spherical covering surface (below). The supercharges can map the state $\phi$ onto $\chi$ lying in a multi-cycle twisted sector, i.e.~a descendant of $\sigma_{(w-k,k)}$. Such a state $\chi$ has the property that its lift to the covering surface is the product of two vertex operators $\hat{\chi}_k$, $\hat{\chi}_{w-k}$ inserted at different points.}\label{fig:cov_surf}
\end{figure}

\subsection{Anomalous dimensions and \texorpdfstring{$\frac{1}{N}$}{1/N} expansion}

Let us now go back to the mixing-matrix $\tilde{\gamma}$ in eq.~(\ref{eq:gamma_decomposition}), and in particular its $\frac{1}{N}$ expansion. We will work to leading order in the coupling constant $g$ --- for the matrix elements of  $\tilde{\gamma}$ this is at order $g^2$ (since each individual $3$-point function is of order $g$) ---  and study its $\frac{1}{N}$ expansion, which is of the form
\begin{equation}
\tilde{\gamma} = \tilde{\gamma}_0  \Bigl( \frac{1}{N} + \frac{d_0}{N^2} + \cdots \Bigr) + \,\tilde{\gamma}_1 \Bigl( \frac{1}{N^2} + \frac{d_1}{N^3} + \cdots \Bigr)\ ,
\end{equation}
where $\tilde{\gamma}_0$ are the leading order terms from the sphere diagrams that were considered in \cite{Gaberdiel:2023lco}, while $\tilde{\gamma}_1$ comes from the transitions in eq.~(\ref{eq:gamma_decomposition}) for which the intermediate states $\ket{\chi}$ are in a $(w-k,k)$ twisted sector with $k=2,\ldots,w-2$.\footnote{There are also $\mcl{O}\bbr{\frac{1}{\sqrt{N}}}$ corrections for $\ket{\chi}$ in the $w-1$ twisted sector where the untwisted colour is excited, and they are also included in $\tilde{\gamma}_1$.}
The non-planar contributions are sketched in Figure \ref{fig:cov_surf}. From the perspective of string theory on AdS$_3$, $g$ controls the amount of R-R flux that is being switched on, while $N$ is related to the string coupling constant via $g_s^2 \sim 1/N$.

We should mention that there are also $\frac{1}{N^2}$ corrections from the sphere diagrams (the $\tilde{\gamma}_0$ term). They arise from expanding out the combinatorical binomial factors appearing in the rewriting of orbifold correlators in terms of covering surfaces, see e.g.\ \cite[eq.~(2.35)]{Pakman:2009zz}. These corrections are the reason why it is often more natural to expand the answer not in $\frac{1}{N}$, but rather in the fugacity $p$ of the grand canonical symmetric orbifold $\bigoplus_N p^N \text{Sym}^N\bbr{\mbb{T}^4}$ \cite{Eberhardt:2020bgq,Aharony:2024fid}. We will comment on the contributions to the $\frac{1}{N^2}$ corrections where appropriate, but will mainly focus on the calculation of $\tilde{\gamma}_1$.

\noindent The planar anomalous dimensions are obtained by diagonalising $\tilde{\gamma}_0$,
\begin{equation}
\tilde{\gamma}_0\ket{\epsilon_0} =  \epsilon_0\,\ket{\epsilon_0}\ .
\end{equation}
The non-planar corrections that we are interested in can then be obtained as an expectation value, just as in standard QM perturbation theory,\footnote{If the eigenvalues are degenerate, one has to diagonalise $\tilde{\gamma}_1$ on the degenerate subspace instead. The low lying eigenvalues we will study in Section \ref{sec:results} are usually non-degenerate.}
\begin{equation}\label{eq:total_anomalous_dimension}
\begin{aligned}
\epsilon &=\frac{1}{N}\,\epsilon_0
+\frac{1}{N^2}\,\bbr{\epsilon_1+d_0\,\epsilon_0}
+\cdots\ ,\\
\epsilon_1&=\bra{\epsilon_0}\tilde{\gamma}_1\ket{\epsilon_0}\ .
\end{aligned}
\end{equation}
The planar calculation was done in quite some detail in \cite{Gaberdiel:2023lco}, and it was found that the result simplifies dramatically for large $w$. In particular, in this limit, the sum over the intermediate states from the sectors associated to $w\pm 1$ can be restricted further to those states that preserve the number of magnon excitations. Indeed, the contribution of states for which the number of magnon modes is not preserved are individually at least $\mcl{O}\bbr{\frac{1}{w}}$ suppressed, and it was checked in \cite{Gaberdiel:2024nge} that they do not modify the result significantly at large $w$ (despite the fact that there are ${\cal O}(w)$ many such terms).

For the states we consider in this paper, the planar eigenvalues $\epsilon_0$ come from intermediate transitions into the $w-1$ sector. The combinatorial prefactor of orbifold correlators for such a three-point function is \cite{Pakman:2009zz}
\begin{equation}
\begin{aligned}
&\sqrt{\frac{(N-w)!(N-w+1)!(N-2)!}
{N!(N-w)!}} \\
&\qquad=\frac{1}{\sqrt{N}}\,
\br{1-\frac{w-2}{2N}+\cdots}\ .
\end{aligned}
\end{equation}
Thus, the coefficient $d_0$ in the $\frac{1}{N}$ expansion of $\gamma_0$ (which is a sum over squares of three-point functions) is given by
\begin{equation}
d_0 = 2-w \ .
\end{equation}

\subsection{Calculational method for correlators}\label{sec:method}

The non-planar corrections to the eigenvalues can be calculated using the transitions, see eq.~(\ref{eq:gamma_decomposition}) and Figure \ref{fig:cov_surf},
\begin{equation}
\bra{\chi}\tilde{Q}_2\ket{\phi}\ ,\qquad \bra{\chi}\tilde{S}_2\ket{\phi}\ ,
\end{equation}
where $\ket{\phi}$ is a state in the $w$-, and $\ket{\chi}$ is a state in the $(w-k,k)$ twisted sector. These matrix elements are short-hand for a regulated three-point function, see \cite{Gaberdiel:2023lco}. For example,
\begin{align}\label{eq:3point_transition}
\bra{\chi}&\tilde{S}_2\ket{\phi} = \nonumber\\
&= \frac{i\pi}{\sqrt{2}}\,\delta_{h_\phi,h_\chi}\nonumber\\
&\quad\times
\corr{V\bbr{\ket{\chi}^\dagger,\infty}\,
V\bbr{G^-_{-\frac{1}{2}}\ket{2},1}
V\bbr{\ket{\phi},0}
}\nonumber\\
&= \frac{-i\pi}{\sqrt{2}}\,\delta_{h_\phi,h_\chi}\nonumber\\
&\quad\times
\corr{
V\bbr{\ket{\phi}^\dagger,\infty}\,
V\bbr{G^+_{\frac{1}{2}}\ket{2}^\dagger,1}
\times V\bbr{\ket{\chi},0}
}^*\ .
\end{align}
In the second line, we have written the matrix element as the complex conjugate of a correlator where the state $\ket{\chi}$ is inserted at 0; this configuration is preferable for computational reasons. Furthermore, $V(\cdot)$ is a vertex operator, and the operator inserted at $1$ comes from the contraction of $\tilde{S}_2$ with the perturbation $\Phi$. Finally, the $\delta_{h_\phi,h_\chi}$ term is a consequence of the regularisation procedure, and it enforces that $\ket{\chi}$ and $\ket{\phi}$ have the same left-moving conformal dimension.

These correlators can be calculated using the covering map method of Lunin and Mathur \cite{Lunin:2000yv}. By the Riemann--Hurwitz formula, one sees that the covering surface in this case is a sphere, and the covering map is given by the polynomial
\begin{equation}\label{eq:cov_map}
\Gamma_{w,k}(z) = \br{\frac{k-w}{k}}^k\,z^{w-k}\,\br{z-\frac{w}{w-k}}^k\ .
\end{equation}
Here, the twist operators $\sigma_{(w-k,k)}$, $\sigma_2$, $\sigma_w$ are inserted at the positions $0$, $1$, and $\infty$, respectively, as in the second line of eq.~(\ref{eq:3point_transition}) above. Note that $0$ has two ramified pre-images at $z=0$ and $z=\frac{w}{w-k}$ with ramification index $w-k$ and $k$, respectively, implementing the multi-cycle twist.

More concretely, the  matrix element can be calculated as follows. The states $\ket{\phi}$, $\ket{\chi}$ can be written as descendants of the BPS states \begin{equation}
\ket{w} \qquad \text{and} \qquad \ket{w-k,k}:=\ket{w-k}\otimes\ket{k}\ ,
\end{equation}
respectively, see Appendix \ref{app:BPS_states}. (Note that $\ket{w-k,k}$ has dimension and charge $\frac{w-2}{2}$.) The descendants we consider will always be expressed in terms of fractional $\mathbb{T}^4$ modes, see Appendix \ref{app:conventions} for our conventions, which can be lifted to ordinary modes of the free fields on the covering surface. For example, if
\begin{equation}
\ket{\chi} = \alpha_{-\frac{m}{w-k}}\ket{\chi'}\ ,
\end{equation}
where $\alpha$ is a boson acting on the $w-k$ cycle, the correlator $\langle \cdots V(\ket{\chi},0)\rangle$ is lifted to the covering surface as \cite{Gaberdiel:2023lco}
\begin{equation}
\begin{aligned}
\llangle \cdots V\bbr{\ket{\chi},0}\rrangle
={}&\cint{C(0)} dz\,
\br{\Gamma_{w,k}(z)}^{-\frac{m}{w-k}}\\
&\quad\times
\llangle \cdots \hat{\alpha}(z)\,
V\bbr{\ket{\hat{\chi}'},0}\rrangle\ .
\end{aligned}
\end{equation}
Here and below the fields on the covering surface are distinguished by a hat. By iterating this lifting of fractional modes, the correlator can finally be written as the contour integral of a correlator of free fields on the covering surface. Note that since the BPS states are not simply the ground states of the respective twisted sectors, they also need to be lifted, but their lift is relatively simple, see e.g.\
\cite{Lunin:2001pw}. We should also mention that depending on whether the fractional descendant of  $\ket{w-k,k}$ acts on the $w-k$ or the $k$-cycle, the contour integral on the covering surface needs to be taken around $z=0$ or $z=\frac{w}{w-k}$, respectively, see eq.~(\ref{eq:cov_map}).

\section{Numerical results}\label{sec:results}

In the following, we will first study this problem numerically, and present (numerical) evidence that non-planar corrections  break the degeneracy that was found in \cite{Gaberdiel:2023lco} at large twist. We will examine two families of states that have degenerate planar eigenvalues at large $w$, and determine how these results are affected by the $\epsilon_1$ corrections. The calculation is computationally intensive for large $w$ since  the number of relevant states increases exponentially with $w$; as a consequence we will only be able to solve the problem exactly for small values of $w$, $w\leq 11$. Nevertheless, already from these calculations one can see that while the difference in the planar dimensions of the two sets of states decreases as $w$ increases (as predicted by \cite{Gaberdiel:2023lco}), the non-planar corrections do not show any such behaviour.

In Section~\ref{sec:degeneracy_breaking} we will study the structure of this $\frac{1}{N}$ calculation more abstractly. As we will explain there, the calculations for the two sets of states are structurally quite different, and there is, {\it a priori}, no obvious reason why their $\frac{1}{N}$ corrections should agree. Together with the fact that the numerical calculations for $w\leq 11$ do not show any sign for such a `miraculous' agreement, we regard this as good evidence for the fact that the degeneracy of these eigenvalues is indeed lifted by the $\frac{1}{N}$ corrections.

\subsection{The explicit results}\label{sec:3.1}

To be more specific, we shall in the following compare the anomalous dimensions of the two families of states
\begin{equation}\label{eq:reference_states}
\begin{gathered}
\alpha^1_{-\frac{m}{w}}
\alpha^1_{-1+\frac{m}{w}}\ket{w}\ ,\\
\psi^-_{\frac{1}{2}-\frac{m}{w}}
\psi^-_{-\frac{1}{2}+\frac{m}{w}}\ket{w}\ ,
\qquad m=1,\dots,w-1\ .
\end{gathered}
\end{equation}
where $\ket{w}$ is the BPS state with dimension and charge $\frac{w-1}{2}$ in the $w$-twisted sector.
In the following, we will refer to these families of states as ``bosonic'' and ``fermionic'', respectively, to distinguish their different constituent magnons. One should note, however, that both have the same bosonic statistics.
In the planar limit and for $w\to\infty$, these states have the same anomalous dimension \cite{Gaberdiel:2023lco}
\begin{equation}\label{eq:3.2}
\frac{g^2}{N}\,\frac{\sin^2\bbr{\pi \frac{m}{w}}}{\frac{m}{w}\bbr{1-\frac{m}{w}}}\ .
\end{equation}
They are thus degenerate at large $w$. While supersymmetry relates individual bosons to fermions and vice versa, the two families in question are not related to one another by the $\mcl{N}=(4,4)$ algebra, and thus there is a priori no algebraic reason for this degeneracy.\footnote{As for the case of $\mcl{N}=4$ SYM, this additional degeneracy is therefore regarded as evidence for integrability.}

\begin{figure*}[ht]
\centering
\begin{minipage}[t]{.45\textwidth}
\begin{tikzpicture}
\node[](n1)at(0,0){\includegraphics[width=.8\linewidth]{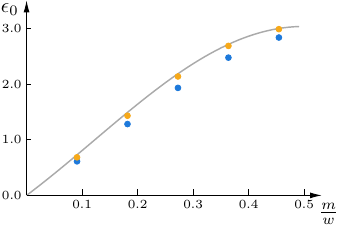}};
\node[anchor=north east]at(n1.north west){\large(a)};
\end{tikzpicture}
\end{minipage}
\hfill
\begin{minipage}[t]{.45\textwidth}
\begin{tikzpicture}
\node[](n1)at(0,0){\includegraphics[width=.8\linewidth]{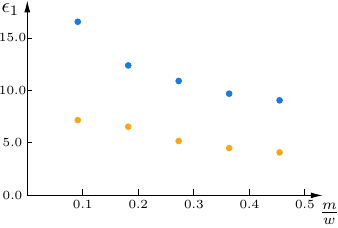}};
\node[anchor=north east]at(n1.north west){\large(b)};
\end{tikzpicture}
\end{minipage}
\caption{The spectra for the bosonic $\alpha^1_{-\frac{m}{w}}\alpha^1_{-1+\frac{m}{w}}$ (blue) and fermionic $\psi^-_{\frac{1}{2}-\frac{m}{w}}
\psi^-_{-\frac{1}{2}+\frac{m}{w}}$ (orange) states for $w=11$. In (a), we show the planar spectrum $\epsilon_0$, which follows the predicted dispersion relation eq.~(\ref{eq:3.2}) (grey line) relatively well. The difference between the spectra is consistent with a $1/w$ correction. Plot (b) shows the non-planar corrections $\epsilon_1$ to the spectrum. They are not in agreement.}\label{fig:spectra_full_calc}
\end{figure*}

We have implemented the full non-planar calculation in Mathematica, which allows one to carry out the complete calculation of anomalous dimensions at small $w$. This includes the complete diagonalisation of the planar problem, including states with higher magnon number, see \cite{Gaberdiel:2024nge} for more details. For the largest value we have studied completely ($w=11$), the results are given in Figure~\ref{fig:spectra_full_calc}. In the left panel we show the planar, i.e.\ $\epsilon_0$ term in the notation of eq.~(\ref{eq:total_anomalous_dimension}), while the non-planar corrections ($\epsilon_1$) are shown on the right.

One observes that the planar spectra are close (consistent with an $\mcl{O}\bbr{\frac{1}{w}}$ difference, see Section~\ref{sec:3.2} below) and well described by the prediction from the magnon-preserving approximation of \cite{Gaberdiel:2023lco}. However, the non-planar spectra differ significantly by terms of the order $\mcl{O}(w)$ for small mode number $\frac{m}{w}$.

We should mention that the planar anomalous dimensions of the bosonic states are smaller than those of the fermionic states, while the non-planar ones are larger. As a consequence, the non-planar anomalous dimensions are different both in the grand canonical ensemble, as well as in an expansion in powers of $\frac{1}{N}$. Indeed, as explained in eq.~(\ref{eq:total_anomalous_dimension}), in the $\frac{1}{N}$ expansion, the planar dimension also contributes at order $\frac{1}{N^2}$, giving a total result of
\begin{equation}
\epsilon_1 + (2-w)\,\epsilon_0\ .
\end{equation}
Since the coefficient $2-w$ is negative, this pushes the non-planar corrections even further apart.

\begin{figure}[t]
\centering
\includegraphics[width=\linewidth]{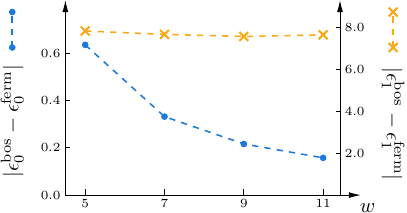}
\caption{The average difference of the anomalous dimensions of bosonic and fermionic states plotted for odd $w$ from $5$ to $11$. Left axis (blue dots): The average difference of the planar spectrum. This difference clearly decreases for increasing $w$, consistent with a $\mcl{O}\bbr{\frac{1}{w}}$ behaviour. Right axis (orange crosses): The average difference of non-planar corrections $\epsilon_1$. This average appears to be largely independent of $w$, and no sign of a restoration of the degeneracy at large $w$ is visible.}\label{fig:average_differences}
\end{figure}

\subsection{The behaviour as a function of $w$}\label{sec:3.2}

Since the degeneracy of the spectrum is only expected to emerge at large $w$, it is important to understand how the $1/N$ corrections behave as a function of $w$. To illustrate this we have plotted the differences in anomalous dimensions as a function of the twist length $w$, see Figure~\ref{fig:average_differences}. As is clear from the plot, the average difference in the planar anomalous dimensions decreases with increasing $w$, and is consistent with a $\mcl{O}\bbr{\frac{1}{w}}$ asymptotic behaviour. The average difference in the non-planar anomalous dimensions, on the other hand, appears to remain more or less independent of $w$.

While these findings are obviously suggestive, the exact results are only available for relatively small values of $w$, and it is therefore important to understand a bit more conceptually the origin of these differences. This will be explained in the next section.

\section{Degeneracy-breaking mechanisms}\label{sec:degeneracy_breaking}

In this section, we discuss the structural differences in the calculation of the non-planar corrections
in some detail. In particular, we will pinpoint a number of qualitative differences between the contributions to the two family of states which persist for large $w$. Together with our explicit small $w$ results this suggest that there is no restoration of the degeneracy as $w\to\infty$.

\subsection{The bosonic states}\label{sec:bosonic_state}

In order to understand the origin of the differences we need to explain the calculation of the two kinds of states in (\ref{eq:reference_states}) in a little more detail. We begin with the (normalised) bosonic states
\begin{equation}\label{phim}
\begin{aligned}
\ket{\phi_m} &:=\frac{1}{\sqrt{m(w-m)}}\,
\alpha^1_{-\frac{m}{w}}
\alpha^1_{-1+\frac{m}{w}}\ket{w}\ ,\\
&\hspace{4em}m=1,\dots,w-1\ .
\end{aligned}
\end{equation}
The planar anomalous dimension $\epsilon_0$ associated to $\ket{\phi_m}$ is obtained by choosing the eigenstate $\ket{\epsilon_0}$ of $\tilde{\gamma}_0$ with the largest overlap with $\ket{\phi_m}$ \cite{Gaberdiel:2024nge},
\begin{equation}\label{4.2}
\ket{\epsilon_0} = c\,\ket{\phi_m} + \dots\ ,
\end{equation}
where the dots denote higher magnon contributions. As explained in Section \ref{sec:method}, the non-planar correction $\epsilon_1$ to the eigenvalue is the expectation value $\bra{\epsilon_0}\tilde{\gamma}_1\ket{\epsilon_0}$, which can be computed as a sum over products of three-point functions
\begin{equation}\label{4.3}
\begin{aligned}
\epsilon_1
&=\bra{\epsilon_0}\tilde{\gamma}_1\ket{\epsilon_0}\\
&=\sum_{\chi:\,\text{multi-cycle}}\Bigl[
\bra{\epsilon_0}\tilde{S}_2\ket{\chi}
\bra{\chi}\tilde{Q}_2\ket{\epsilon_0}\\
&\hspace{5em}+
\bra{\epsilon_0}\tilde{Q}_2\ket{\chi}
\bra{\chi}\tilde{S}_2\ket{\epsilon_0}\Bigr]\ .
\end{aligned}
\end{equation}
Here, the sum is over intermediate states in a multi-cycle twisted sector, giving contributions which are subleading in the $\frac{1}{N}$ expansion.

It is natural to expect that the dominant contribution to $\epsilon_1$ comes from the expectation value $\bra{\phi_m}\tilde{\gamma}_1\ket{\phi_m}$ with respect to the leading term in eq.~(\ref{4.2}), i.e.\ (\ref{phim}).
In the following  we will analyse which intermediate states $\chi$ gives rise to the dominant contributions for this matrix element $\bra{\phi_m}\tilde{\gamma}_1\ket{\phi_m}$, see eq.~(\ref{4.3}); in the next subsection we will compare these results to the analogous analysis for the fermionic states. The qualitative differences we will find suggest that these non-planar corrections will continue to differ between the bosonic and the fermionic states even at large twist $w$.

The intermediate states $\ket{\chi}$ contributing most to $\bra{\phi_m}\tilde{\gamma}_1\ket{\phi_m}$ are those with the same number of magnons (i.e.~fractional $\mbb{T}^4$ modes) acting on the BPS vacuum. Under $\tilde{Q}_2$, the state $\ket{\phi_m}$ is mapped to
\begin{equation}
\ket{\chi} = \alpha^1_{-q}\psi^+_{-\frac{3}{2}+q}\ket{w-k,k}\ ,
\end{equation}
where $q$ is either in $\frac{1}{w-k}\mbb{Z}$ or $\frac{1}{k}\mbb{Z}$, depending on whether the mode acts on the $w-k$- or $k$-cycle. For the special case that $q$ can be written both as $q = \frac{n}{w-k}$ and as $q = \frac{n'}{k}$, both modes can act on either cycle. This can only happen if $\text{gcd}(w-k,k)>1$, or if $q \in \mbb{Z}$. For example, for $q=1$, there are always the two different configurations corresponding to
\begin{equation}
\begin{gathered}
\Bigl(\alpha^1_{-1}\psi^+_{-\frac{1}{2}}\ket{w-k}\Bigr)
\otimes\ket{k}\ ,\\
\Bigl(\alpha^1_{-1}\ket{w-k}\Bigr)
\otimes\Bigl(\psi^+_{-\frac{1}{2}}\ket{k}\Bigr)\ .
\end{gathered}
\end{equation}
as well as those where $k$ and $w-k$ are interchanged, and we need to include all of these terms. It turns out that these (half-)integer mode number states give actually a large contribution to the anomalous dimension.

Under the action of $\tilde{S}_2$, there are, on the other hand, no magnon-number preserving transitions. This is due to the global $\mfr{su}(2)$ symmetry rotating the left- and right-moving bosons, which is preserved by the perturbation, see Appendix \ref{app:conserved_charges}. To satisfy this charge conservation constraint, the state $\ket{\chi}$ therefore needs to include at least three bosonic excitations.\footnote{One can also see this directly by writing out the states and the perturbing field explicitly, and then working out the Wick contractions of the constituent fundamental fields.}

\begin{figure*}[t]
\centering
\begin{tikzpicture}[scale=0.9]
\node[](n1)at(0,0){$\alpha^1\alpha^1\ket{w}$};
\node[](n1pQ)[above left=0.75cm and 0.5cm of n1]{$\alpha^1\psi^+\ket{w-1}$};
\node[](n1pS)[above right=0.75cm and 0.5cm of n1]{$0$};
\draw[-{Latex[length=2.5mm,width=1mm]}](n1.north west)--(n1pQ.south east)[xshift=0.5em,yshift=0.5em]node[midway]{$\tilde{Q}_2$};
\draw[-{Latex[length=2.5mm,width=1mm]}](n1.north east)--(n1pS.south west)[xshift=-0.5em,yshift=0.5em]node[midway]{$\tilde{S}_2$};
\node[align=center](n1nQ)[below left=0.75cm and 0.25cm of n1]{$\alpha^1\psi^+\ket{w-k,k}$\\ + $\alpha^1_{-1}\psi^+_{-\frac{1}{2}}\ket{w-k,k}$\\ + 3-magnon (6)};
\node[](n1nS)[below right=0.75cm and 0.5cm of n1]{\hspace{-1.5cm}3-magnon (2)};
\draw[-{Latex[length=2.5mm,width=1mm]}](n1.south west)--(n1nQ.north east)[xshift=-0.6em,yshift=0.5em]node[midway]{$\tilde{Q}_2$};
\draw[-{Latex[length=2.5mm,width=1mm]}](n1.south east)--(n1nS.north west)[xshift=0.5em,yshift=0.5em]node[midway]{$\tilde{S}_2$};

\node[](n2)at(7.1,0){$\psi^-\psi^-\ket{w}$};
\node[](n2pQ)[above left=0.75cm and 0.5cm of n2]{$\alpha^2\psi^-\ket{w-1}$};
\node[](n2pS)[above right=0.75cm and 0.5cm of n2]{$0$};
\draw[-{Latex[length=2.5mm,width=1mm]}](n2.north west)--(n2pQ.south east)[xshift=0.5em,yshift=0.5em]node[midway]{$\tilde{Q}_2$};
\draw[-{Latex[length=2.5mm,width=1mm]}](n2.north east)--(n2pS.south west)[xshift=-0.5em,yshift=0.5em]node[midway]{$\tilde{S}_2$};
\node[align=center](n2nQ)[below left=0.75cm and 0.00cm of n2]{$\alpha^2\psi^-\ket{w-k,k}$\\ + 3-magnon (3)};
\node[align=center](n2nS)[below right=0.75cm and 0.5cm of n2]{\hspace{-0.5cm}$\alpha^1\psi^-\ket{w-k,k}'$\\ \hspace{-1em}+ 3-magnon (3)};
\draw[-{Latex[length=2.5mm,width=1mm]}](n2.south west)--([xshift=-1em]n2nQ.north east)[xshift=-0.6em,yshift=0.5em]node[midway]{$\tilde{Q}_2$};
\draw[-{Latex[length=2.5mm,width=1mm]}](n2.south east)--(n2nS.north west)[xshift=0.5em,yshift=0.5em]node[midway]{$\tilde{S}_2$};

\node[]at(3.5,0.5){$\big\uparrow$ \textbf{planar} $\big\uparrow$};
\node[]at(3.5,-0.5){$\big\downarrow$ \textbf{non-planar} $\big\downarrow$};
\draw[dashed,gray](1.5,0)--(5.5,0);
\end{tikzpicture}
\caption{Sketch of the dominant transitions from the fermionic and bosonic states at large $w$. In the top half we describe the planar transitions, for which the two states behave alike. The bottom half gives a schematic description of the non-planar transitions, for which the two states behave differently. First, for the fermionic states, there are magnon-number preserving transitions under $\tilde{S}_2$. (Here, $\ket{w-k,k}'$ schematically stands for a BPS state with the correct charges, as explained in the text.) Second, the bosonic state can be mapped to $\alpha^1_{-1}\psi^+_{-\frac{1}{2}}\ket{w-k,k}$ under $\tilde{Q}_2$, which collectively makes a large difference. Lastly, ``3-magnon ($n$)'' denotes the number $n$ of classes of intermediate three-magnon states, which sum to give a large contribution. There are a different number of such classes for the bosons and fermions.}\label{fig:transition_sketch}
\end{figure*}
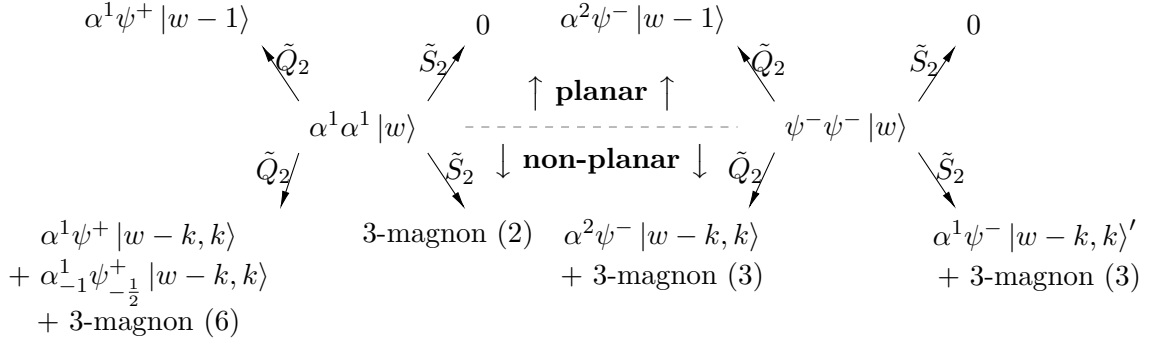

However, this does not mean that we shouldn't include any contributions from the second term in (\ref{4.3}). Indeed, when inspecting the transitions that actually contribute  to the non-planar corrections for $w\leq 11$, we find that there are some non-magnon-number preserving states that give a large contribution.\footnote{This happens both for the intermediate states under the action of $\tilde{Q}_2$, i.e.\ the first term in eq.~(\ref{4.3}), as well as for the intermediate states under the action of $\tilde{S}_2$, i.e.\ the second term in eq.~(\ref{4.3}).} These contributions arise from intermediate four-magnon states which contain a $\psi^+_{-1/2}$ or $\bar{\psi}^+_{-1/2}$ mode. For example, under $\tilde{Q}_2$, there are relevant transitions to
\begin{equation}\label{4.6}
\alpha^1_{-q_1}\alpha^1_{-q_2}\alpha^2_{-1+q_1+q_2}\bar{\psi}^+_{-\frac{1}{2}}\ket{w-k,k}\ .
\end{equation}
Since $\bar{\psi}^+_{-1/2}\ket{w-k,k}$ is again a BPS state, it is a bit a matter of taste whether we think of this state as a 4-magnon state, or a 3-magnon excitation relative to a different BPS vacuum. This `explains' in a sense why the transitions to the states in eq.~(\ref{4.6}) are more important than those that involve generic intermediate
 4-magnon states. We have found that in order to approximate the answer for $w\leq 11$ well, we need to include certain such `3-magnon' states, and they are described explicitly in Appendix~\ref{app:three_magnon_states}, see the 8 intermediate families of states described in eqs.~(\ref{A.10}) and (\ref{A.11}). At $w=9$, $11$, we find that intermediate 2- and 3-magnon states capture approximately $75\%$, $80\,\%$ of the contributions to $\bra{\phi_m}\tilde{\gamma}_1\ket{\phi_m}$, respectively. We therefore believe that these intermediate states play the dominant role also at large $w$. The resulting situation (for the bosonic states) is sketched in the left panel of Figure~\ref{fig:transition_sketch}.

We should note that it is not surprising that transitions to states with higher magnon number play an important role in our non-planar analysis. Recall from \cite[Section 3.2]{Gaberdiel:2024nge} that magnon-number violating transitions are suppressed by powers of the cycle length. In the planar case, transitions occur only to magnon excitations in the $w\pm 1$ sectors, and thus these transitions are all small at large $w$. For the non-planar corrections, on the other hand, the intermediate states contain cycles of length $w$ and $w-k$, where $k$ runs over all possible values, $k=2,\ldots, w-2$. Thus for a given $w$, the intermediate magnons will also act on cycles of length $k\ll w$, and as a consequence, transitions to higher magnon states are only suppressed by $\mcl{O}\bbr{\frac{1}{k}}$. They can therefore give rise to a non-negligible contribution to the supercharge action.

\subsection{The fermionic states}\label{sec:fermionic_state}

We now turn to the second family of states under consideration,
\begin{equation}
\begin{aligned}
\ket{\psi_m} &:=\frac{1}{w}\,
\psi^-_{\frac{1}{2}-\frac{m}{w}}
\psi^-_{-\frac{1}{2}+\frac{m}{w}}\ket{w}\ ,\\
&\hspace{4em}m=1,\dots,w-1\ .
\end{aligned}
\end{equation}
Again, we describe the dominant contributions to the expectation value $\bra{\psi_m}\tilde{\gamma}_1\ket{\psi_m}$ that determines the non-planar correction. As we shall see there  are important qualitative differences in how the supercharges $\tilde{Q}_2$ and $\tilde{S}_2$ act on these states, compared to the bosonic analysis of the previous section.

First, it follows from the charge conservation discussion of Appendix \ref{app:conserved_charges} that $\tilde{Q}_2$ can map this state to
\begin{equation}\label{3.12}
\ket{\chi} = \alpha^2_{-q}\psi^-_{-\frac{1}{2}+q}\ket{w-k,k}\ .
\end{equation}
This should be compared to the corresponding transition for the bosonic states
\begin{equation}
\tilde{Q}_2:\,\alpha^1_{-\frac{m}{w}}\alpha^1_{-1+\frac{m}{w}}\ket{w} \longmapsto \alpha^1_{-q}\psi^+_{-\frac{3}{2}+q}\ket{w-k,k}\ .
\end{equation}
While the individual contractions making up this transition are similar, see Appendix \ref{app:contractions}, the range of $q$ is in fact not exactly the same. In particular, for the state
\be
\alpha^1_{-q}\psi^+_{-\frac{3}{2}+q}\ket{w-k,k}\ ,
\ee
the mode number $q=1$ is allowed, whereas this is not the case for the right-hand-side of (\ref{3.12})
due to the relation\footnote{Note that if we had worked with the top BPS state as a reference state instead, there would be a similar effect with the $\psi^+_{-\frac{1}{2}}$ mode.}
\begin{equation}\label{3.161}
\psi^-_{\frac{1}{2}}\ket{w-k,k} = 0\ .
\end{equation}
Thus, in the sum over $k$, there is a consistent difference in the number of transitions. These turn out to have a cumulative effect for the states with small $m$.

Second, unlike the discussion for the bosonic states, $\tilde{S}_2$ can also map the state $\ket{\psi_m} $ to a magnon-number perserving state in the $(w-k,k)$-twisted sector, namely to
\begin{equation}\label{3.16}
\alpha^1_{-q}\psi^-_{-\frac{1}{2}+q}\ket{w-k,k}_\mrm{L}\otimes \ket{\widetilde{\text{BPS}}}_\mrm{R}\ ,
\end{equation}
where we have explicitly separated left- and right-movers, and $\ket{\widetilde{\text{BPS}}}_\mrm{R}$ stands schematically for a right-moving BPS state with dimension and R-charge $\frac{w}{2}$. We list these states and the values of the relevant three-point functions in Appendix \ref{app:BPS_states}. It is instructive to compare this to the planar calculation, where this effect does not appear at leading order in $w$. Specifically, in the planar calculation, the analogous transitions under $\tilde{S}_2$ are from the $w$- to the $(w-1)$-twisted sector. Thus, there are $w$ active colours in the correlator, and the four right-moving BPS states with the correct charges in the $(w-1)$-twisted sector (together with one untwisted colour), are\footnote{There are four such BPS states with allowed charges in the $(w-k,k)$-twisted sector, see Appendix \ref{app:BPS_states}. Their contributions have to be summed.}
\begin{align}\label{newstates}
\ket{\widetilde{\text{BPS}}}_\mrm{R}
&=\widetilde{\ket{w-1}}_+\otimes\widetilde{\ket{0}}\ ,\nonumber\\
\ket{\widetilde{\text{BPS}}}_\mrm{R}
&=\widetilde{\ket{w-1}}_-\otimes
\widetilde{\bbr{K^+_{-1}\ket{0}}}\ ,\nonumber\\
\ket{\widetilde{\text{BPS}}}_\mrm{R}
&=\widetilde{\ket{w-1}}_{0_+}\otimes
\widetilde{\bbr{\bar{\psi}^+_{-\frac{1}{2}}\ket{0}}}\ ,\nonumber\\
\ket{\widetilde{\text{BPS}}}_\mrm{R}
&=\widetilde{\ket{w-1}}_{0_-}\otimes
\widetilde{\bbr{\psi^+_{-\frac{1}{2}}\ket{0}}}\ .
\end{align}
The transition to the first state is of order $\mcl{O}\bbr{\frac{1}{w}}$, and hence does not contribute in the large twist limit. Naively, the transition to the second state is $\mcl{O}(1)$, but the state is actually multi-particle, because the second factor is not in the vacuum. Taking the proper orbifold averaging into account \cite{Pakman:2009zz}, one finds that the transition is $\mcl{O}\bbr{\frac{1}{\sqrt{N}}}$ suppressed and hence does not contribute to the planar calculation of \cite{Gaberdiel:2023lco}. However, this state \emph{does} contribute at first subleading order in $1/N$. Finally, the transitions to the last two states are both $\mcl{O}\bbr{\frac{1}{\sqrt{w N}}}$ suppressed. Thus in the planar limit $\tilde{S}_2=0$, see the top line of Figure~\ref{fig:transition_sketch}, and this mirrors therefore the behaviour of the bosons in the planar limit. However, unlike the situation for the bosons, there is now also a magnon-number perserving state in the non-planar limit (from the second state in (\ref{newstates})). In addition, there are again certain 3-magnon intermediate states which can give large contributions: for the fermionic states at $w=11$, the 2- and 3-magnon transitions make up almost $95\,\%$ of the $\bra{\psi_m}\tilde{\gamma}_1\ket{\psi_m}$ expectation value. The 3-magnon states are listed in Appendix~\ref{app:three_magnon_states}, see the six families of states in eq.~(\ref{A.12}) and eq.~(\ref{A.13}).

The individual transitions behave quite similarly to the bosonic states in (\ref{A.10}) and (\ref{A.11}), and they can be calculated using the contractions listed in Appendix~\ref{app:contractions}. At large $w$, these are approximated by simple trigonometric functions as in \cite{Gaberdiel:2023lco}, and the expressions for the contractions for bosonic and fermionic states agree. (At finite $w$, there are $\mcl{O}\bbr{\frac{1}{w}}$ differences between the contractions.) This similarity in the individual transitions essentially leads to the degeneracy of the planar spectrum at large $w$. However, the qualitative differences in the dominant transitions outlined above and summarised in Figure \ref{fig:transition_sketch} still lead to differences in the full evaluation of the non-planar corrections. For example, analysing the contributions of 3-magnon transitions, we have found that for $w\leq 11$ the effect of the 3-magnon states is roughly twice as large for the bosonic states compared to the fermionic states.

While the details are quite complicated (and we do not have a simple explanation for some of the qualitative behaviours), it should be clear from this discussion that the calculation for the bosonic and fermionic states are, on the face of it, quite different. While this does not prove that the degeneration is lifted in the large $w$ limit, it would require a rather special conspiracy to prevent this from happening --- and our numerical calculations of the previous section do not seem to suggest that anything of this kind actually takes place. Taken together we regard this as strong evidence for the fact that integrability is broken by these $\frac{1}{N}$ effects.

\section{Conclusion}\label{sec:conclusion}

In this paper, we have studied the anomalous dimensions of two archetypal families of states in the symmetric product orbifold, deformed by the modulus dual to turning on R-R flux in string theory. The two classes of states are not related to one another by supersymmetry, but their planar anomalous dimensions agree in the limit of large twist $w$. We have calculated the non-planar $\frac{1}{N}$ corrections to the anomalous dimensions and we have found strong evidence that the degeneracy is lifted by these corrections.

The calculation involves evaluating transitions to a large class of intermediate states in multi-cycle twisted sectors. Individually, these transitions behave similarly to those studied in \cite{Gaberdiel:2023lco}, however, their large number enhances $\mcl{O}\bbr{\frac{1}{w}}$ differences to lift the planar large $w$ degeneracy. Furthermore, the effect of higher-magnon corrections is more relevant than in the planar calculation, and affects the two families of states differently. As a consequence, there is no structural reason why the degeneracy should persist also to order $1/N$, and our explicit numerical calculations for small $w$ ($w\leq 11$), see Section~\ref{sec:results}, do not show any signs that this in fact happens.

Non-planar corrections have also been studied in other instances of the AdS/CFT correspondence,  in particular for $\mathcal{N}=4$ SYM in \cite{Beisert:2003tq}, and for ABJM in \cite{Kristjansen:2008ib}, and our results show that the $\text{AdS}_3/\text{CFT}_2$ duality behaves similarly in this regard. However, our setup deviates in some important structural ways from these two examples. First of all,  the orbifold theory is dual to an $\text{AdS}_3$ background with pure NS-NS flux  \cite{Gaberdiel:2018rqv, Eberhardt:2018ouy,Eberhardt:2019ywk}, whereas the higher dimensional examples all involve pure R-R flux. 
In addition, the pairs of states with degenerate planar dimensions studied in~\cite{Beisert:2003tq, Kristjansen:2008ib} are distinguished by their parity, while their field content is the same. As our theory is vector-like, i.e.~the fundamental fields in our theory are not matrix-valued fields, there is no analogous notion of a trace, and hence no natural parity operation; we have therefore studied two families of states that involve different magnon excitations. 
Finally, the integrable structure of the orbifold at large $N$ has so far not been understood as a spin chain, and integrability is more subtle since it only appears manifestly in the large $w$ limit. (At finite $w$, there are probably large wrapping corrections.)

In the companion paper \cite{GNP2} we have also studied the distribution of level spacings, and we have seen further qualitative differences between the planar and non-planar spectrum. In particular, the planar level spacings are sampled from a Poisson distribution, while the non-planar corrections show clear signs of level repulsion. Taken together this therefore gives strong evidence for the claim that the deformation of the orbifold is only integrable at large $N$, and that the $\frac{1}{N}$ corrections break this integrability.

\section*{Acknowledgements}

We thank Yiming Chen, Frank Coronado, Bin Guo, Ji Hoon Lee, Anthony Houppe,  Juan Maldacena, Edward Mazenc, Kiarash Naderi, Vit Sriprachyakul, Haoyu Zhang for useful conversations, and Haoyu Zhang for helpful comments on the draft. MRG thanks the Institute for Advanced Study, Princeton, for hospitality during the final stages of this work; his stay there was supported by the Adler Family Fund. The work of BN is supported through a personal grant of MRG from the Swiss National Science Foundation. The work of the group at ETH is also supported in part by the Simons Foundation grant 994306 (Simons Collaboration on Confinement and QCD Strings), as well the NCCR SwissMAP that is also funded by the Swiss National Science Foundation. CP is supported by NSFC NO.~12675092,, 12175237, NO.~12447108, and in part by NO.~12247103, the Fundamental Research Funds for the Central Universities, and funds from the Chinese Academy of Sciences.

\appendix

\section{Conventions}\label{app:conventions}

\subsection{Field content of \texorpdfstring{$\mbb{T}^4$}{T4}}

We denote the four left-moving bosons and fermions of $\mbb{T}^4$ by $\alpha^i,\bar{\alpha}^i$, $i=1,2$ and $\psi^\pm,\bar{\psi}^\pm$, respectively; they satisfy the OPE relations
\begin{equation}
\bar{\alpha}^i(x)\alpha^j(y)\sim \frac{\epsilon^{ij}}{\big(x-y\big)^2}\ ,\qquad \bar{\psi}^\pm(x)\psi^\mp(y)\sim \frac{\pm 1}{x-y} \ .
\end{equation}
The right-moving fields will always be distinguished by a tilde. The (left-moving) $\mcl{N}=4$ generators are built out of these fields as
\begin{equation}\label{N4fields}
\begin{aligned}
G^+&=\bar{\alpha}^2\,\psi^+
+\alpha^2\,\bar{\psi}^+\ ,\\
G'^+&=-\bar{\alpha}^1\,\psi^+
-\alpha^1\,\bar{\psi}^+\ ,\\
G^-&=\bar{\alpha}^1\,\psi^-
+\alpha^1\,\bar{\psi}^-\ ,\\
G'^-&=\bar{\alpha}^2\,\psi^-
+\alpha^2\,\bar{\psi}^-\ ,\\
K^+&=\bar{\psi}^+\,\psi^+\ ,\qquad
K^-=-\bar{\psi}^-\,\psi^-\ ,\\
K^3&=\frac{1}{2}:
\bar{\psi}^+\,\psi^-+\bar{\psi}^-\,\psi^+:\ .
\end{aligned}
\end{equation}
and
\begin{equation}
\begin{aligned}
L={}&:\bar{\alpha}^1\,\alpha^2
-\bar{\alpha}^2\,\alpha^1:\\
&+\frac{1}{2}:\psi^+\,\partial\bar{\psi}^-
+\bar{\psi}^-\,\partial\psi^+\\
&\hspace{2.5em}-\bar{\psi}^+\,\partial\psi^-
-\psi^-\,\partial\bar{\psi}^+:\ .
\end{aligned}
\end{equation}
These fields generate the ${\cal N}=4$ superconformal algebra with central charge $c=6$. Throughout the text we use the short-hand notation
\begin{equation}
\tilde{S}_2 = \tilde{G}'^+_{-\frac{1}{2}}\ ,\qquad \hbox{and} \qquad \tilde{Q}_2 = \tilde{G}'^-_{\frac{1}{2}}\ .
\end{equation}

\subsection{Conserved charges}\label{app:conserved_charges}

The perturbation preserves the following charges, see also \cite{Gaberdiel:2024nge}. Firstly, it  commutes with the $\mfr{su}(2)$ $R$-symmetry of the left- and right-movers independently, so both $K^3_0$ and $\tilde{K}^3_0$ are preserved. Secondly, it follows from the analysis in \cite[Section 3.4]{Gaberdiel:2023lco}, that to first order, a state can only transition under the action of $\tilde{S}_2$ or $\tilde{Q}_2$ to a state with the same (free) left-moving dimension $L_0$. Thirdly, there is a residual $\mfr{su}(2)$ symmetry rotating the bosons, and this leads to a residual charge $U^\text{L/R}_\text{res}$ for both the left- and right-movers, under which the $\alpha^1,\bar{\alpha}^1$ bosons are charged positively and the $\alpha^2,\bar{\alpha}^2$ bosons are charged negatively. The perturbation preserves the diagonal charge
\begin{equation}
U_\text{res} := U^\text{L}_\text{res} + U^\text{R}_\text{res}\ .
\end{equation}
In particular, the supercharges $\tilde{S}_2$ and $\tilde{Q}_2$ are charged positively and negatively under $U_\text{res}$, respectively. And finally, there is a $\mfr{u}(1)$ symmetry with generator $U_\text{bar}$, under which unbarred modes are charged positively and barred modes are charged negatively. The perturbation preserves the bar charge of the left- and right-movers independently.

\subsection{BPS states and background correlators}\label{app:BPS_states}

In the $w$ twisted sector, there are four left-moving BPS states $\ket{w}_a$, which are $\psi^+_{-1/2}$ and $\bar{\psi}^+_{-1/2}$ descendants of the lowest BPS state $\ket{w}_-$ \cite{Lunin:2001pw}. They have the following charges and labels $a$:
\begin{center}
\begin{tabular}{c|c|c|c|c}
$a$ & $L_0$ & $K^3_0$ & $U_\text{bar}$ & $U_\text{res}$\\ \hline &&&\\[-8pt]
$-$ & $\frac{w-1}{2}$ & $\frac{w-1}{2}$ & $0$ & $0$ \\[5pt]
$0_+$ & $\frac{w}{2}$ & $\frac{w}{2}$ & $1$ & $0$ \\[5pt]
$0_-$ & $\frac{w}{2}$ & $\frac{w}{2}$ & $-1$ & $0$ \\[5pt]
$+$ & $\frac{w+1}{2}$ & $\frac{w+1}{2}$ & $0$ & $0$
\end{tabular}
\end{center}
If the subscript $a$ is suppressed, we mean the lowest BPS state $a=-$ by convention. For example,
\begin{equation}
\ket{w-k,k} := \ket{w}_-\otimes \ket{k}_-\ .
\end{equation}

In the calculation of the transition elements, there is a factor coming from the `background correlator' of the reference BPS states. These can be calculated as in \cite{Lunin:2001pw}, and we list the results for completeness. For $\tilde{Q}_2$ transitions, the background correlator is
\begin{equation}\label{eq:q2_bg}
\bra{w}_-\, \sigma_2(1)\ket{w-k}_-\otimes \ket{k}_- = \sqrt{\frac{w}{(w-k)\,k}}\ ,
\end{equation}
while for $\tilde{S}_2$ transitions, there are four transitions
\begin{equation}
\bra{w}_-\,\sigma_2^\dagger(1)\,\ket{w-k}_a\otimes\ket{k}_b
\end{equation}
allowed by charge conservation (such that the state in the $(w-k,k)$ sector has dimension and charge $\frac{w}{2}$), and the background correlators are
\begin{align}\label{eq:s2_bg}
\bra{w}_-\,\sigma_2^\dagger(1)\,\ket{w-k}_+\otimes\ket{k}_- & =
\sqrt{\frac{k}{w\,(w-k)}}\ ,\nonumber\\
\bra{w}_-\,\sigma_2^\dagger(1)\,\ket{w-k}_-\otimes\ket{k}_+ & =
\sqrt{\frac{w-k}{w\,k}}\ ,\nonumber\\
\bra{w}_-\,\sigma_2^\dagger(1)\,\ket{w-k}_{0_+}\otimes\ket{k}_{0_-} & =
\sqrt{\frac{1}{w}}\ ,\nonumber\\
\bra{w}_-\,\sigma_2^\dagger(1)\,\ket{w-k}_{0_-}\otimes\ket{k}_{0_+} & =
\sqrt{\frac{1}{w}}\ .
\end{align}
Importantly, the sum of the squares of eq.~(\ref{eq:s2_bg}) add up to the square of eq.~(\ref{eq:q2_bg}). This means that in the calculation of $\tilde{\gamma}_1$, $\tilde{S}_2$ and $\tilde{Q}_2$ effectively have the same background correlator, as the left- and right-movers are independent.
Alternatively, one can also see that $\tilde{S}_2$ and $\tilde{Q}_2$ give rise to the same contribution using that $\sigma_2^\dagger = K^-_{1}\sigma_2$, writing the $K^-_{1}$ mode in terms of a contour integral, and then using familiar contour wrapping arguments.

\subsection{Additional intermediate states}{\label{app:three_magnon_states}}

As discussed in Section \ref{sec:bosonic_state}, higher magnon corrections are important for the non-planar anomalous dimension of the bosonic states. Here, we write down the families of intermediate `3-magnon' states which give a large contribution to the non-planar anomalous dimension, both for the bosonic and fermionic states. These states are 4-magnon descendants of $\ket{w-k,k}$ which contain at least one $\psi^+_{-\frac{1}{2}}$ or $\bar{\psi}^+_{-\frac{1}{2}}$ mode.\footnote{Importantly, the $\psi^+_{-\frac{1}{2}}$ mode can act on either the $w-k$ or $k$ cycle.} They can thus be interpreted as a 3-magnon excitation of a different BPS state.

Under $\tilde{Q}_2$, there are transitions from $\alpha^1\alpha^1\ket{w}$ to
\begin{align}
&\alpha^1_{-q_1}\alpha^1_{-q_2}\bar\alpha^2_{-q_3}\psi^+_{-\frac{1}{2}}\ket{w-k,k}\ ,\nonumber\\
& \alpha^1_{-q_1}\bar\alpha^1_{-q_2}\alpha^2_{-q_3}\psi^+_{-\frac{1}{2}}\ket{w-k,k}\ ,\nonumber\\
& \alpha^1_{-q_1}\bar\psi^+_{-q_2}\psi^-_{-q_3}\psi^+_{-\frac{1}{2}}\ket{w-k,k}\ , \label{A.10}\\
& \alpha^1_{-q_1}\psi^+_{-q_2}\psi^-_{-q_3}\bar\psi^+_{-\frac{1}{2}}\ket{w-k,k}\ ,\nonumber\\
& \alpha^1_{-q_1}\psi^+_{-q_2}\bar\psi^-_{-q_3}\psi^+_{-\frac{1}{2}}\ket{w-k,k}\ ,\nonumber\\
& \alpha^1_{-q_1}\alpha^1_{-q_2}\alpha^2_{-q_3}\bar\psi^+_{-\frac{1}{2}}\ket{w-k,k}\ . \nonumber
\end{align}
The states in the third and fourth line come from the same four-magnon state, but the $-\frac{1}{2}$ mode is on a different fermion. Under $\tilde{S}_2$, there are fewer transitions,
\begin{align}
&\alpha^1_{-q_1}\alpha^1_{-q_2}\alpha^1_{-q_3}\bar\psi^+_{-\frac{1}{2}}\ket{w-k,k}'\ ,\nonumber\\
&\alpha^1_{-q_1}\alpha^1_{-q_2}\bar\alpha^1_{-q_3}\psi^+_{-\frac{1}{2}}\ket{w-k,k}'\ , \label{A.11}
\end{align}
where $\ket{w-k,k}'=\ket{w-k,k}_\text{L}\otimes \ket{\widetilde{\text{BPS}}}_\text{R}$ is a BPS state with the correct charges as discussed in Appendix \ref{app:BPS_states}.

Similarly, for $\psi^-\psi^-\ket{w}$, $\tilde{Q}_2$ can map onto the following three-magnon states,
\begin{align}
&\alpha^2_{-q_1}\psi^-_{-q_2}\bar\psi^-_{-q_3}\psi^+_{-\frac{1}{2}}\ket{w-k,k} \ ,\nonumber\\
&\alpha^2_{-q_1}\psi^-_{-q_2}\psi^-_{-q_3}\bar\psi^+_{-\frac{1}{2}}\ket{w-k,k} \ , \label{A.12}\\
&\bar\alpha^2_{-q_1}\psi^-_{-q_2}\psi^-_{-q_3}\psi^+_{-\frac{1}{2}}\ket{w-k,k} \ .\nonumber
\end{align}
The other supercharge $\tilde{S}_2$ behaves analogously, and can map onto
\begin{align}
&\alpha^1_{-q_1}\psi^-_{-q_2}\bar\psi^-_{-q_3}\psi^+_{-\frac{1}{2}}\ket{w-k,k}' \ ,\nonumber\\
&\alpha^1_{-q_1}\psi^-_{-q_2}\psi^-_{-q_3}\bar\psi^+_{-\frac{1}{2}}\ket{w-k,k}' \ , \label{A.13}\\
& \bar\alpha^1_{-q_1}\psi^-_{-q_2}\psi^-_{-q_3}\psi^+_{-\frac{1}{2}}\ket{w-k,k}'\ . \nonumber
\end{align}
Note that there are thus eight families of intermediate three-magnon states for $\alpha^1\alpha^1\ket{w}$, while there are only six for $\psi^-\psi^-\ket{w}$.

\onecolumngrid
\section{Contractions}\label{app:contractions}

The transitions into the $(w-k,k)$ sector can be calculated using a few building blocks coming from Wick contractions on the covering surface. Firstly, there are `same-species' contractions, which come from contractions of constituent magnons of the external states. Secondly, due to the presence of the perturbation $\Phi$, there are `crossed' contractions, where a boson transitions to a fermion or vice versa \cite{Gaberdiel:2023lco}.

We denote the same-species contractions by a bracket above the magnons and crossed contractions by a crossed-out bracket below the magnons. These are short-hand notations for the corresponding  three-point functions. Concretely,
\begin{align}
\sscontraction{\bar{\alpha}^2_{1-\frac{m}{w}}}{\alpha^1_{-1+\frac{n}{w-k}}}
&\equiv \tfrac{1}{\sqrt{(w-k-n)\,(w-m)}}\,\bra{w}\bar{\alpha}^2_{1-\frac{m}{w}}\,V\bbr{\ket{2},1}\,\alpha^1_{-1+\frac{n}{w-k}}\ket{w-k,k}\ ,\nonumber\\
\ccontraction{\bar{\alpha}^2_{1-\frac{m}{w}}}{\psi^+_{-\frac{3}{2}+\frac{n}{w-k}}}
&\equiv \tfrac{1}{\sqrt{(w-k)\,(w-m)}}\,\bra{w}\bar{\alpha}^2_{1-\frac{m}{w}}\,V\bbr{G'^+_{\frac{1}{2}}\ket{2},1}\,\psi^+_{-\frac{3}{2}+\frac{n}{w-k}}\ket{w-k,k}\ ,
\end{align}
and similarly for the other transitions. The pre-factors come from the normalisation of the modes.

The contractions are evaluated using the method of \cite{Lunin:2000yv} outlined in Section \ref{sec:method}. The explicit form of the covering map is given in eq.~(\ref{eq:cov_map}). As an example, the contractions needed for the evaluation of the anomalous dimension of $\alpha^2\alpha^2\ket{w}$ in the magnon-number preserving approximation are given below.
The same-species contraction in this case is
\begin{align}
\sscontraction{\bar{\alpha}^2_{1-\frac{m}{w}}}{\alpha^1_{-1+\frac{n}{w-k}}}
&= \frac{(-1)^k \left(\frac{w-k}{k}\right)^{-k\frac{m}{w}}  \left(\frac{w}{k}\right)^{k\frac{n}{w-k}}    \left(\frac{w}{k-w}\right)^{n-m}}{w\,k\,(w-k) \sqrt{(w-m) (w-k-n)}} \nonumber \\
&\quad\times ((k-w) (w-m)+w)\left(-w (k+n+1)+k+w^2\right)\nonumber\\
&\quad\times \frac{1}{\frac{n}{w-k}-\frac{m}{w}}\,\binom{k \bbr{-1+\frac{n}{w-k}}}{w-k-n-1}\binom{k\bbr{1-\frac{m}{w}}}{w-m-1} \ .
\end{align}
In the last line, there is the mode number-conservation favouring factor $\frac{1}{\frac{n}{w-k}-\frac{m}{w}}$, familiar from \cite{Gaberdiel:2023lco}.

On the other hand, the crossed contraction with the fermion acting on the $w-k$ cycle is given by
\begin{align}
\ccontraction{\bar{\alpha}^2_{1-\frac{m}{w}}}{\psi^+_{-\frac{3}{2}+\frac{n}{w-k}}}
&= \sqrt{w-m}\,\frac{w}{k}\,\bbr{-\tfrac{w}{w-k}}^{w-m-1}\,\bbr{\tfrac{w-k}{k}}^{k(1-\frac{m}{w})}\,\binom{k\bbr{1-\frac{m}{w}}}{w-m}\nonumber\\
&\quad \times (-1)^{w-k-n} \bbr{\tfrac{w-k}{w}}^{(w-k)-n + k(1-\frac{n}{w-k})+\frac{1}{2}}\,\bbr{\tfrac{w-k}{k}}^{k(-\frac{3}{2}+\frac{n}{w-k})}\nonumber\\
&\quad \times \binom{k\bbr{-1+\frac{n}{w-k}}-1}{w-k-n}\ .
\end{align}
The contractions for the $\psi^-\psi^-\ket{w}$ state are of a similar form: for example the same-species contractions are related by
\begin{align}
\sscontraction{\bar{\psi}^+_{\frac{1}{2}-\frac{m}{w}}}{\psi^-_{-\frac{1}{2}+\frac{n}{w-k}}}
&= \frac{(w-k)(w-m-1)}{w(w-k-n-1)+k}
\sqrt{\frac{1-\frac{n}{w-k}}{1-\frac{m}{w}}}
\,\sscontraction{\bar{\alpha}^2_{1-\frac{m}{w}}}{\alpha^1_{-1+\frac{n}{w-k}}}\ .
\end{align}
For the full calculation in Section \ref{sec:results}, additional contractions are needed. These come from Wick contractions where fields from the same external state contract among each other, see also \cite{Gaberdiel:2024nge}, and can again be expressed in terms of products of binomial coefficients.

\newpage
\twocolumngrid

\end{document}